\documentclass[journal]{IEEEtran}

\usepackage[pdftex]{graphicx}
\usepackage{amsmath}
\usepackage{amssymb}
\usepackage{url}
\usepackage{xcolor}
\usepackage[export]{adjustbox}
\graphicspath{{images/}}
\begin{document}
	\title{Investigating Shift-Variance of Convolutional Neural Networks in Ultrasound Image Segmentation}
	
	\author{Mostafa~Sharifzadeh,
		Habib~Benali,~\IEEEmembership{Fellow,~IEEE,}
		and~Hassan~Rivaz,~\IEEEmembership{Senior~Member,~IEEE}
		\thanks{M. Sharifzadeh, H. Benali, and H. Rivaz are with the Department
			of Electrical and Computer Engineering, Concordia University, Montreal, QC, Canada, and with PERFORM Center, Concordia University, Montreal, QC, Canada.}
		}
	\maketitle
	
	\begin{abstract}
		While accuracy is an evident criterion for ultrasound image segmentation, output consistency across different tests is equally crucial for tracking changes in regions of interest in applications such as monitoring the patients’ response to treatment, measuring the progression or regression of the disease, reaching a diagnosis, or treatment planning. Convolutional neural networks (CNNs) have attracted rapidly growing interest in automatic ultrasound image segmentation recently. However, CNNs are not shift-equivariant, meaning that if the input translates, e.g., in the lateral direction by one pixel, the output segmentation may drastically change. To the best of our knowledge, this problem has not been studied in ultrasound image segmentation or even more broadly in ultrasound images. Herein, we investigate and quantify the shift-variance problem of CNNs in this application and further evaluate the performance of a recently published technique, called BlurPooling, for addressing the problem. In addition, we propose the Pyramidal BlurPooling method that outperforms BlurPooling in both output consistency and segmentation accuracy. Finally, we demonstrate that data augmentation is not a replacement for the proposed method. Source code is available at \url{https://git.io/pbpunet} and \url{http://code.sonography.ai}.
	\end{abstract}
	
	
	\section{Introduction}
	\label{sec:Introduction}
	Image segmentation entails pixel-level labelling of images to obtain a representation of data that is more meaningful and easier to analyze for a specific purpose. Ultrasound image segmentation is a crucial task in numerous applications such as registration \cite{Hu2012}, image-guided biopsy and therapy \cite{Hricak2007}, automatic staging of stenoses in intravascular ultrasound (IVUS) diagnosis \cite{ Katouzian2012}, detection of vessel boundaries to monitor cardiovascular diseases \cite{Park2020}, and delineation of the cardiac structures \cite{Leclerc2019, Painchaud2020a}.
	
	Although manual segmentation is regarded as the gold standard in many applications, it is tedious and impossible to perform fast enough for real-time applications. Automatic segmentation has been a highly sought-after research area in medical image processing to overcome these issues. While traditional image processing techniques such as graph cuts and active contours had been state-of-the-art (SOTA) for a long time, approaches based on CNNs have attracted increasing attention recently.
	
	\begin{figure}
		\centering
		\includegraphics[width=0.9999\linewidth]{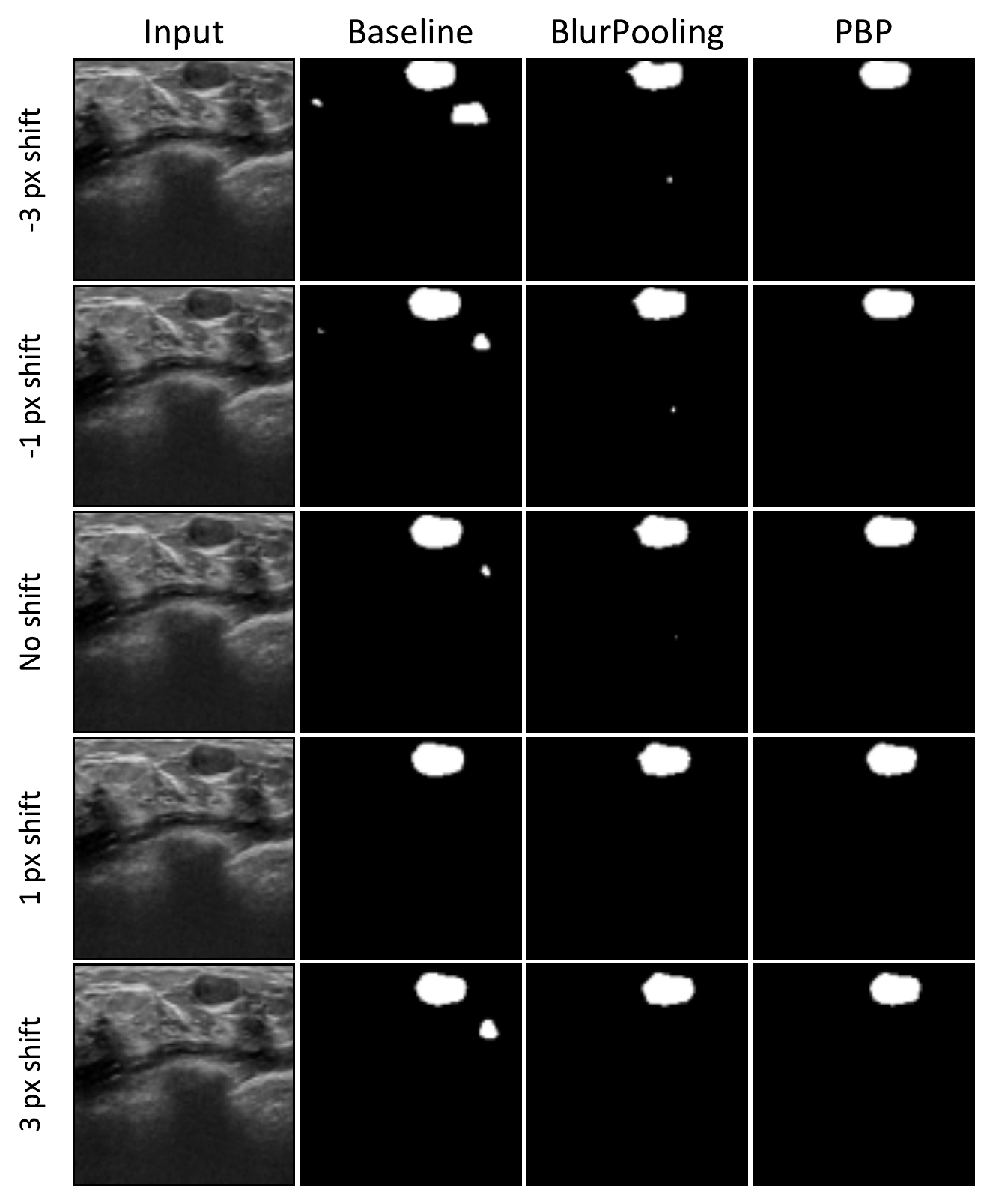}
		\caption{Illustration of the effect of input translations on generating output segmentation masks. The left column shows an identical input image from the test set translated diagonally for $k$ pixels, where $k \in \{-3, -1, 0, +1, +3\}$. The following columns show corresponding segmentation masks generated by the baseline method, BlurPooling ($7\times 7$), and the proposed method (Pyramidal BlurPooling), respectively.}
		\label{fig1}
		\vspace{-15pt}
	\end{figure}
	
	However, while CNNs had been considered shift-invariant for many years, it has been recently reported that they are subject to the shift-variance problem \cite{Engstrom2019d, Zhang2019, Kayhan2020, Azulay2019a}, \cite{Lee2020, Chaman2021}. In other words, if the input translates by only one pixel, the segmentation result may change.
	Although this drawback may be tolerated in applications such as natural image classification, it hinders CNNs' performance in sensitive applications, such as medical image segmentation, where the reliance of the networks on main features, as well as reproducibility of the results, are essential. Since in many scenarios, changes in images are being tracked to monitor the progression or regression of the disease or the patients’ response to the therapy. In addition, in image-guided interventions wherein the target moves with breathing or other physiological motions, shift-equivariance is paramount. All of this brings us to the conclusion that in addition to accuracy, consistency is also a crucial metric that needs to be considered while evaluating CNN-based methods for medical applications such as ultrasound image segmentation.
	
	To shed light on the shift-variance problem in CNNs, Fig. \ref{fig1} shows the effect of input translations on output segmentation masks. The left column shows an identical input image from the test set, which was translated diagonally from top-left to bottom-right for -3, -1, 1, and 3 pixels. The next column shows the corresponding segmentation masks generated by a baseline method without addressing the shift-variance problem. The baseline method was affected by small translations and generated substantially different output masks.
	
	\subsection{Related Work}
	\label{ssec:Introduction:RelatedWork}
	CNNs approaches for image segmentation have been extensively investigated in the medical ultrasound community.
	Yap \textit{et al.} \cite{Yap2018a} compared the performance of three deep learning approaches against four traditional SOTA algorithms for breast lesion detection: Radial Gradient Index \cite{Drukker2002}, Multifractal Filtering \cite{Yap2008}, Rule-based Region Ranking \cite{Shan2012}, and Deformable Part Models \cite{Pons2013}. Besides, to overcome the lack of public datasets in this domain, they made a breast lesion ultrasound dataset available for research purposes.
	In 2D echocardiographic images, Smistad \textit{et al.} developed a multi-view network to segment the left ventricular in different apical views \cite{Smistad2019a}, and Leclerc \textit{et al.} evaluated several encoder-decoder CNNs for segmenting cardiac structures and estimating clinical indices \cite{Leclerc2019}. They also developed the Refining U-Net (RU-Net) and a multi-task Localization U-Net (LU-Net) to refine and improve the robustness of segmentation \cite{Leclerc2019c, Leclerc2020, Leclerc2019b}.
	Abraham \textit{et al.} addressed the data imbalance issue \cite{Sudre2017} in lesion segmentation by proposing a generalized focal loss function based on the Tversky index and combining it with an improved version of an attention U-Net \cite{Abraham2019}.
	Karimi \textit{et al.} proposed three different methods to estimate Hausdorff Distance from the segmentation probability map produced by a CNN, and suggested three loss functions for training CNNs that lead to a reduction in Hausdorff Distance without degrading other segmentation metrics such as the Dice similarity coefficient \cite{Karimi2020}.
	Gu \textit{et al.} introduced a comprehensive attention-based CNN (CA-Net) by making extensive use of multiple attentions in a CNN architecture for more accurate and explainable medical image segmentation. They claimed that the network is aware of the most important spatial positions, as well as channels and scales at the same time \cite{Gu2021}.
	For prostate segmentation in 2D and 3D transrectal ultrasound images, van Sloun \textit{et al.} employed a variant of U-net \cite{VanSloun2018, VanSloun2019, VanSloun2021}, and Wang \textit{et al.} developed a 3D deep neural network equipped with attention modules by harnessing the deep attentive features \cite{Wang2019}.
	Li \textit{et al.} employed three modified U-Nets combined with the concept of cascaded networks in IVUS images \cite{Li2021}.
	Looney \textit{et al.} presented a multi-class CNN for real-time segmentation of the placenta, amniotic fluid, and fetus in 3D ultrasound \cite{Looney2021}. For a similar purpose, Zimmer \textit{et al.} used an auxiliary task to improve the performance and introduced a method to extract the whole placenta at late gestation using multi-view images \cite{Zimmer2019, Zimmer2020}.
	Zhou \textit{et al.} proposed a fully automated solution to segment the myotendinous junction region in successive ultrasound images in a single shot using a region-adaptive network (RAN), which learns about the salient information of the myotendinous junction \cite{Zhou2020}.
	They also introduced an approach that combined a voxel-based fully convolution network (Voxel-FCN) and a continuous max-flow post-processing module to automatically segment the carotid media-adventitia (MAB) and lumen-intima boundaries (LIB) and to generate the vessel-wall-volume (VWV) measurement from three-dimensional ultrasound images \cite{Zhou2020a}.
	Park \textit{et al.} proposed a technique to improve the measurement accuracy of the flow velocity in arteries, especially in the near-wall region, by introducing a U-Net-based architecture called USUNet followed by compensation for the effect of wall motion \cite{Park2020}. Amiri \textit{et al.} exploited test time augmentation to improve the accuracy of segmentation of ultrasound images \cite{Amiri2020a}. They also investigated several different transfer learning schemes in ultrasound segmentation \cite{Amiri2020b}.
	
	\subsection{Contributions}
	\label{ssec:Introduction:Contributions}
	While the output robustness against input translations can be crucial in medical applications, in the literature, the main focus has primarily been on improving segmentation accuracy metrics such as the mean boundary distance, Hausdorff Distance, Dice similarity coefficient (DSC), and volume difference or overlap. Since CNN-based methods are prone to the shift-variance problem, the effect of input translations on the output should not be overlooked. This paper is an extended version of work published in \cite{Sharifzadeh2021}, which to the best of our knowledge, was the first study investigating the shift-variance problem of CNNs in ultrasound image segmentation or even more broadly in ultrasound images. Our contributions are summarized below:
	\begin{itemize}
		\item We investigate the shift-variance problem of CNNs in ultrasound image segmentation using synthetic and \textit{in-vivo} datasets.
		\item A recently published technique, called BlurPooling \cite{Zhang2019}, is applied to mitigate the shift-variance problem, and its performance is evaluated on all datasets.
		\item We propose a new approach, called Pyramidal BlurPooling, which takes into account the nature of the ultrasound segmentation task, and outperforms BlurPooling in both shift-equivariance and segmentation accuracy.
		\item We demonstrate that data augmentation by random translations, a common practice to aid training, is not a replacement for the proposed method.
	\end{itemize}
	
	\section{Background}
	\label{sec:Background}
	\subsection{Shift-variance in CNNs}
	\label{ssec:Background:ShiftVarianceInCNNs}
	One of the motivations for proposing CNN and applying pooling layers as a part of their architectures was making networks robust to irrelevant changes, such as different scales of the same object or image translations \cite{Zeiler2014a}. This idea had been considered valid for many years based on two reasons: the convolutional nature of layers is shift-equivariance, and data augmentation can be employed by feeding the network with variant versions of the same input.
	Utilizing convolutional layers in today's CNNs is inspired by Neocognitron architecture proposed by Fukushima \textit{et al.} \cite{Fukushima1980a} and popularized by LeCun  \textit{et al.} \cite{LeCun1989}. In the Neocognitron architecture, the authors assumed that because all layers are convolutional, the output of the final layer will not be affected by input translations. The data augmentation is motivated by the fact that, for instance, if we randomly crop the input image, the network will see translated versions of the same object during training. Consequently, the trained network will be invariant to both input translations and the absolute spatial location of objects. Nevertheless, it has been reported in the literature that most of the SOTA CNNs are not robust to input translations \cite{Engstrom2019d, Zhang2019, Kayhan2020, Azulay2019a}. Azulay \textit{et al.} showed that the chance of generating a different output by a CNN after translating its input by only a single pixel could be as high as 30\% \cite{Azulay2019a}. 
	
	Neither data augmentation nor convolutional layers guarantee shift-equivariance. If all layers of a CNN were purely convolutional, input translations would be preserved through all layers then, and  the network would be shift-equivariance. However, most modern CNNs contain other layers as well, among which are downsampling layers. It has been suggested that ignoring the Nyquist–Shannon sampling theorem by these downsampling layers is an origin of the shift-variance problem in modern CNNs \cite{Azulay2019a, Zhang2019}. The downsampling operation is mostly performed using pooling or strided-convolutional layers with a stride of more than one. These layers have been employed frequently in commonly-used CNNs such as ResNet \cite{He2016}, VGG \cite{Simonyan2015}, MobileNetV2\cite{Sandler2018}, U-Net \cite{Ronneberger2015}.
	Similarly, showing translated versions of inputs to the network as a data augmentation method may help the network to learn input translations, but  shift-equivariance will be learned merely for similar samples that have been seen before during the training phase. However, the distribution of the training set can be highly biased, and those samples that do not necessarily follow that bias will not take advantage of the learned shift-equivariance. The problem can even be amplified in most medical applications with limited acquired data, where the bias is higher due to the limited number of samples in training sets.

	\subsection{BlurPooling}
	\label{ssec:Background:BlurPooling}
	Consider a band-limited signal, which contains no frequencies higher than \textbf{\textit{B}} Hz. In order to sample the signal without losing any information, the well-known Nyquist–Shannon sampling theorem in the digital signal processing domain indicates that the sampling rate must be higher than \textbf{\textit{2B}} Hz, or otherwise, the reconstructed signal will suffer from aliasing artifact. As mentioned previously, downsampling layers such as max-pooling and strided-convolutional layers do not necessarily respect the Nyquist–Shannon sampling theorem, which leads to sensitivity to input translations and consequently the shift-variance problem \cite{Zhang2019, Azulay2019a}.
	
	A well-known signal processing approach for anti-aliasing is applying a low-pass filter before downsampling. Influenced by this idea, Zhang \textit{et al.} \cite{Zhang2019} proposed merging a low-pass filter with pooling or strided convolutional layers to mitigate the aliasing effect, with minimal additional computation. They referred to the proposed approach as BlurPooling. Although it has been thought that there is a trade-off between blurred-downsampling and max-pooling \cite{Lenc2019}, they demonstrated that they are compatible. This method decomposes pooling and strided-convolutional layers into two separate operations:
	\begin{enumerate}
		\setlength\itemsep{.5em}
		\item Reducing the operation's stride to 1. It is equivalent to apply the same operation densely. For instance, for a max-pooling layer with a 2$\times$2 kernel and stride 2, the max() operator will be applied as previously, but with stride 1 instead of 2. This dense operation preserves shift-equivariance.
		\item Applying an anti-aliasing filter with an $m\times m$ kernel and finally subsampling with the desired factor. Unlike the previous step, which does not change the input dimension, this step reduces the input dimensions as expected.
	\end{enumerate}
	
	\begin{figure}
		\centering
		\includegraphics[width=0.9999\linewidth]{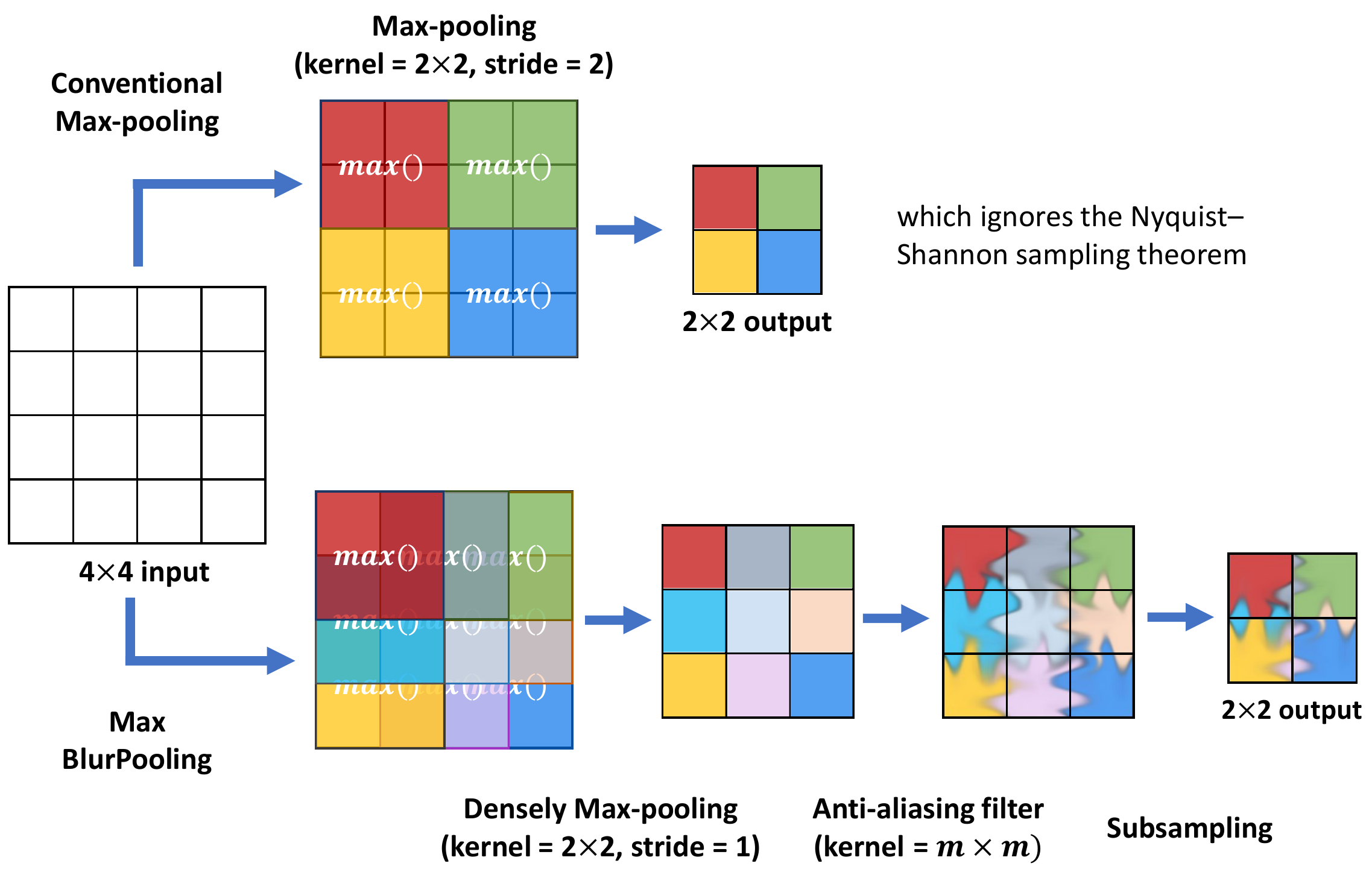}
		\caption{Illustration of the difference between a conventional max-pooling layer and its equivalent BlurPooling layer. (Top path) A conventional max-pooling layer. It does not respect the Nyquist–Shannon sampling theorem during downsampling leading to aliasing artifact and, consequently, lack of shift-equivariance. (Bottom path) A blurpooling layer. It decomposes the max-pool operator into two steps: 1) A densely-evaluated max-pooling. 2) Applying an anti-aliasing filter followed by a subsampling operation.}
		\label{fig2}
	\end{figure}
	
	\noindent Fig. \ref{fig2} illustrates the difference between the implementation of a conventional max-pooling layer and its equivalent BlurPooling (anti-aliased max-pooling) layer. The top path shows a conventional max-pooling layer, which does not respect the Nyquist–Shannon sampling theorem during downsampling leading to aliasing artifact and, consequently, lack of shift-equivariance. The bottom path decomposes this procedure into two steps: 1) A densely-evaluated max-pooling. 2) Applying an anti-aliasing filter followed by a subsampling operation. In the second step, applying the anti-aliasing filter before the subsampling step mitigates the aliasing effect without compromising the advantages of the max() operation. The same concept is also applicable to other downsampling layers, such as strided convolutional layers.
	\noindent This approach suggests augmenting pooling and strided-convolutional layers by low-pass filtering instead of replacing them with averaging operators, which provides shift-equivariance while preserving the advantages of those layers.
	
	\section{Methodology}
	\label{sec:Methodology}
	Let $\boldsymbol{I}\in \mathbb{R}^d$ and $\boldsymbol{\hat{S}}\in \{0,1\}^d$ denote a sample input image and the corresponding output segmentation mask, respectively. The segmentation problem can be formulated as
	\begin{flalign}
	\boldsymbol{\hat{S}} = f_{seg}(\boldsymbol{I}, \boldsymbol{\theta})
	\end{flalign}
	\noindent where $f_{seg}: \mathbb{R}^d \rightarrow \{0,1\}^d$ is the segmentation CNN, and $\boldsymbol{\theta}$ are the network's parameters. By training the CNN, an optimizer is utilized to find optimal parameters $\boldsymbol{\theta^{\ast}}$ that minimize the error, measured by a loss function $L$, between predicted mask $\boldsymbol{\hat{S}}$ and ground truth $\boldsymbol{S}$
	\begin{flalign}
	\boldsymbol{\theta^{\ast}} = \underset{\theta}{argmin} \; L(\boldsymbol{S}, \boldsymbol{\hat{S}})
	\end{flalign}
	
	\subsection{Network architecture}
	The U-Net \cite{Ronneberger2015} was designed for biomedical segmentation applications to work with a limited number of training samples more efficiently. Showing promising results, it is no wonder that the network attracted growing interest in the ultrasound image segmentation domain, and an extended range of networks has been built upon that.
	
	As the primary goal of this study was to investigate the effect of input translations in ultrasound images, we chose the vanilla U-Net as the baseline method to cover an extensive range of previous works. It allows generalizing conclusions of this study and makes them applicable to other work that utilized the original or extended versions of the U-Net without considering the shift-variance problem. Findings can also be valid for studies employing any other networks with conventional downsampling layers, such as max-pooling or strided-convolution, as the source of the problem.
	
	The U-Net consists of a contracting path (encoder) followed by an expansive path (decoder), in which the former extracts locality features, and the latter resamples the image maps with contextual information. Skip connections are also employed to produce more semantically meaningful outputs by concatenating low- and high-level features. Without losing generality, we replaced the transposed convolutions in the original U-Net with bilinear upsampling layers in favor of memory efficiency. For investigating the shift-variance problem, we employed three variants of U-Net by altering its downsampling layers as the source of shift-variance. Fig. \ref{fig3} illustrates the network architectures with three different types of downsampling layers in the contracting path. In the first version, referred to as the baseline network, we utilized max-pooling layers similar to the original U-Net. In the second version, referred to as the BlurPooling network, we replaced the original max-pooling layers with BlurPooling layers, where anti-aliasing filters in all four layers have an identical size of $m\times m$. The third version is described in the following subsection.
	
	\subsection{Pyramidal BlurPooling (PBP)}
	\label{PBP}
	BlurPooling \cite{Zhang2019} was originally introduced for mitigating the shift-variance problem and mainly tested in classification applications and on datasets of natural images, such as CIFAR10 \cite{Krizhevsky2009}, and ImageNet \cite{Russakovsky2015}. In such cases, the predicted probability of the correct class may drop by applying an anti-aliasing filter to feature maps while it climbs, for instance, from the top-3 values to the top-1 value, which leads to preserving or improving the top-N accuracy. However, a concern with adding anti-aliasing filtering is undermining the accuracy of generated segmentation masks, which is critical in sensitive applications, such as medical image segmentation. It has been reported that employing BlurPooling layers in image-to-image translation networks compromises the accuracy \cite{Zhang2019}. It is shown that while the quality of the generated image holds for small anti-aliasing filters, there is a trade-off between quality and shift-invariance for larger ones.
	
\begin{figure}
		\centering
		\includegraphics[width=\linewidth]{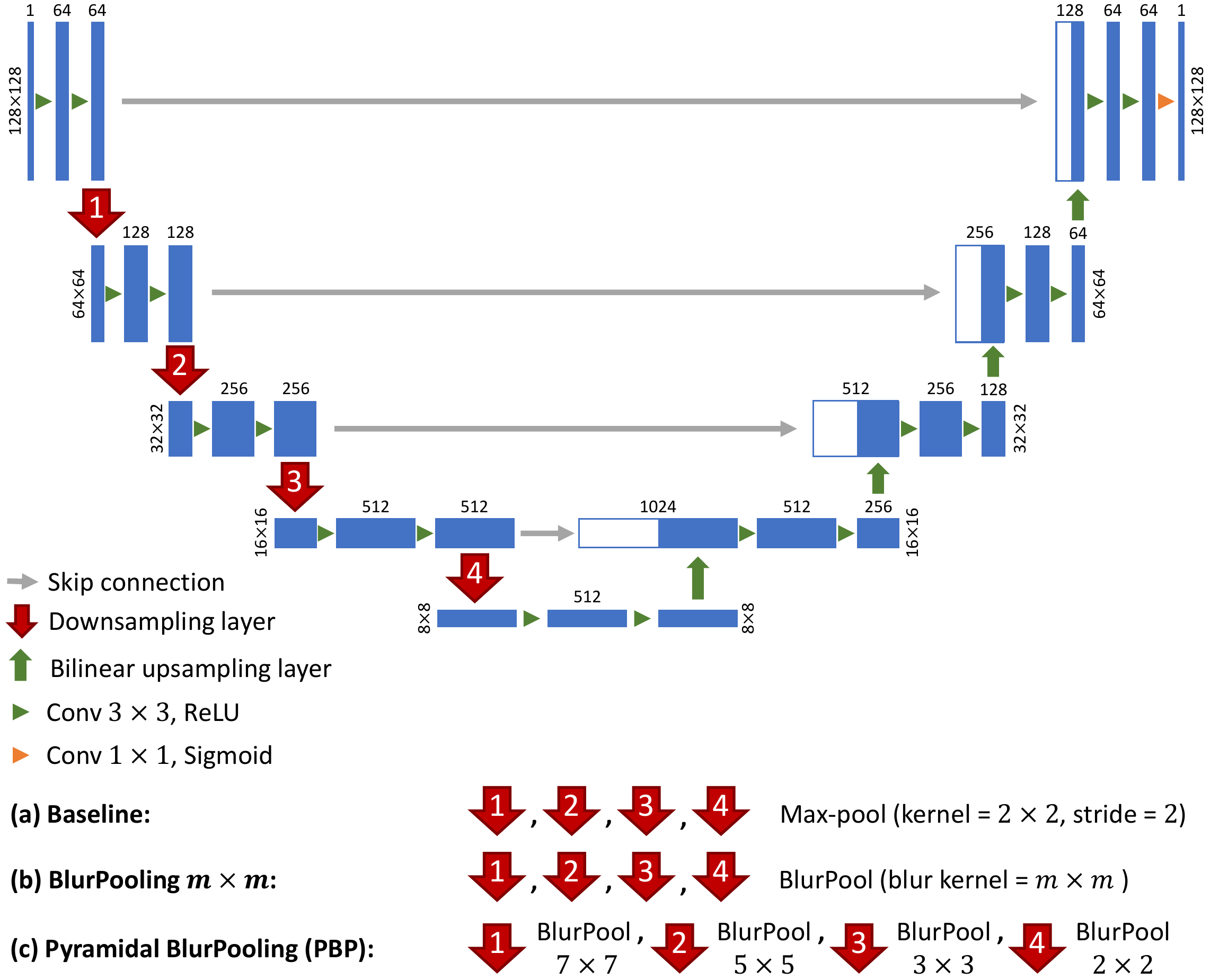}
		\caption{Networks architectures. (a) Similar to the vanilla U-Net, max-pooling layers (kernel=2$\times$2, stride=2) were utilized as downsampling layers. (b) Max-pooling layers were altered with their corresponding BlurPooling layers, in which the size of anti-aliasing filters was identical ($m \times m$) for all four layers. (c) Similar to the previous case, BlurPooling layers were utilized instead of max-pooling layers; however, the size of anti-aliasing filters gradually decreased at each downsampling layer from the first to the fourth one.}
		\label{fig3}
	\end{figure}

	To address this trade-off, we proposed using a pyramidal stack of anti-aliasing filters. As shown in Fig. \ref{fig3}, filters were applied to densely max-pooled feature maps of sizes 128$\times$128$\times c$, 64$\times$64$\times c$, 32$\times$32$\times c$, and 16$\times$16$\times c$ from the first to the fourth downsampling layer, respectively, where $c$ is the number of channels. In Fig. \ref{fig3} (c), labeled as PBP, instead of using anti-aliasing filters of the same size for all downsampling layers, we started with a filter of size 7$\times$7 and made it smaller at each downsampling layer. Moving from the first layer to the last one, more energy of the feature maps concentrates at lower frequencies. Therefore, the shift-equivariance effect can be preserved with a smaller anti-aliasing filter at deeper layers without compromising accuracy.
	
	\subsection{Anti-aliasing filters}
	In this study, we utilized similar smoothing filters used in Laplacian pyramids \cite{Zhang2019, Burt1981, Burt1983}. Let $\boldsymbol{H_s}$ and $\boldsymbol{w_s}$ denote the $m \times m$ filter, and a vector of size $m \times 1$, respectively
	\begin{flalign}
	\boldsymbol{H_s}=\boldsymbol{w_s}\otimes\boldsymbol{w_s}
	\end{flalign}
	
	\noindent where operator $\otimes$ represents the outer product. For the filter $m=2$, we simply chose $\boldsymbol{w_s}=\frac{1}{2}[1, 1]$. For filters $m=2k+1$, $\boldsymbol{w_s}$ was chosen subject to the following constraints \cite{Burt1981}:
	\begin{flalign}
	\text{Normalization:\;\;} \sum_{n=-(m-1)/2}^{(m-1)/2}\boldsymbol{w_s[}n\boldsymbol{]} = 1
	\end{flalign}
	\vspace{-10pt}
	\begin{flalign}
	\text{Symmetry:\;\;} \boldsymbol{w_s[}n\boldsymbol{]} = \boldsymbol{w_s[}-n\boldsymbol{]} \;\;\; for\:all\:n
	\end{flalign}
	\vspace{-15pt}
	\begin{flalign}
	\text{Unimodal:\;\;}
	\boldsymbol{w_s[}n_1\boldsymbol{]} \geq \boldsymbol{w_s[}n_2\boldsymbol{]} \geq 0 \;\;\; for \; 0 \leq n_1 \leq n_2
	\end{flalign}
	\vspace{-15pt}
	\begin{flalign}
	\text{Equal Contribution:\;\;}\sum_{|n|\;even}\boldsymbol{w_s[}n\boldsymbol{]} = \sum_{|n|\;odd}\boldsymbol{w_s[}n\boldsymbol{]}
	\end{flalign}
	
	\noindent In summary, we obtained anti-aliasing filters by taking the outer product of the following vectors with themselves:
	\begin{flalign}
	\boldsymbol{w_s} =
	\begin{cases} 
	\frac{1}{2}[1, 1], & m=2\\
	\frac{1}{4}[1, 2, 1], & m=3\\
	\frac{1}{16}[1, 4, 6, 4, 1], & m=5\\
	\frac{1}{64}[1, 6, 15, 20, 15, 6, 1], & m=7\\
	\end{cases}
	\end{flalign}
	
	\subsection{Datasets}
	\label{ssec:Datasets}
	\subsubsection{Synthetic dataset}
	\label{ssec:Datasets:Synthetic}
	We simulated a synthetic dataset as a reference wherein ground truths are error-free and independent of radiologists' bias to quantify the shift-variance problem in such a scenario. 163 ultrasound images were simulated using the publicly available Field II simulation package \cite{Jensen1996, Jensen1992} containing 100,000 scatterers uniformly distributed inside a phantom of size 50 mm $\times$ 10 mm $\times$ 50 mm in $x$, $y$, and $z$ directions, respectively. All phantoms were positioned at an axial depth of 30 mm from the face of the transducer, and each contained an anechoic region with a different shape. To generate those anechoic regions, instead of using arbitrary shapes, we took corresponding ground truth masks of 163 images of the UDIAT ultrasound breast dataset \cite{Yap2018a} and resampled them with the same size as the phantom. Then we assigned a zero amplitude to scatterers that were located inside the mask. In each simulation, we set the transmit focus at the center of mass of the mask. Finally, we considered masks as the exact ground truths of the simulated images. We split simulated images into three training, validation, and test subsets, each containing 100, 30, and 33 samples, respectively.
	The simulation parameters are summarized in Table \ref{tbl1}.

	\begin{table}[h]
		\caption{Field II parameters for the synthetic dataset.}
		\label{tbl1}
		\setlength{\tabcolsep}{8pt}
		\def\arraystretch{1}%
		\begin{tabular}{p{115pt}p{100pt}}
			\hline
			\vspace{2pt}
			\textbf{Parameter}&
			\vspace{2pt}
			\textbf{Value}
			\vspace{2pt}\\
			\hline
			\vspace{0pt}
			Speed of Sound&
			\vspace{0pt}
			1540 m/s\\
			Center Frequency&
			8.5 MHz\\
			Subaperture Size& 
			64 elements\\
			Number of Scan Lines& 
			100\\
			Element Height& 
			5 mm\\
			Element Width& 
			Equals to wavelength\\
			Kerf& 
			0.05 mm\vspace{2pt}\\
			\hline
		\end{tabular}
	\end{table}
	
	\subsubsection{UDIAT dataset}
	We used a publicly available ultrasound breast images dataset, collected from the UDIAT Diagnostic Centre in 2012 with a Siemens ACUSON Sequoia C512 system and a 17L5 HD linear array transducer (8.5 MHz) \cite{Yap2018a}. It is comprised of 163 breast B-mode ultrasound images, with a mean image size of 760$\times$570 pixels, containing lesions of different sizes at different locations. Lesions were categorized into two classes: benign lesions and cancerous masses, with 110 and 53 samples, respectively. Most of the lesions in the dataset were hypoechoic regions, in which the intensity of the lesion was lower than its background. The dataset also contained respective delineations of the breast lesions as ground truths in separate files, which had been obtained manually by experienced radiologists. We split images into three training, validation, and test subsets, each containing 100, 30, and 33 samples, respectively. Besides, in each subset, we approximately preserved the original ratio of samples per class.
	
	\subsubsection{Baheya dataset}
	To investigate how increasing the number of training samples may affect the shift-variance problem in different scenarios such as with or without augmentation, we employed a larger publicly available dataset collected at Baheya hospital with a LOGIQ E9 ultrasound system equipped with an ML6-15-D Matrix linear probe (1-5 MHz) \cite{Al-Dhabyani2020}. It contained 780 breast ultrasound images with an average image size of 500$\times$500 pixels, collected from 600 female patients with ages ranging from 25 to 75 years old. The dataset was categorized into three classes: normal, benign, and malignant each with 133, 437, 210 cases, respectively. The ground truths, delineated manually by expert radiologists, were presented along with original images. In the case of images containing more than one lesion, each lesion's ground truth had been stored in a separate file. Therefore as a preprocessing step, in those cases, we merged ground truths to have one ground truth file per image. Besides, as the main focus of this work was to investigate the effect of input translations on the predicted segmentation mask, we did not utilize images of the normal class and split the remaining 647 images into three subsets: training, validation, and test, each containing 387, 130, and 130 samples, respectively. Although we reproduced the original classes' ratio in the subsets, we could not separate samples at the subject level because the required information was not available. However, since we fixed the same subsets for all experiments, it did not affect the comparison purposes.
	
	\subsubsection{Mixed ultrasound dataset}
	\label{ssec:Datasets:Mixed}
	To explore the effect of combining two datasets on output consistency, we concatenated training, validation, and test subsets of the UDIAT dataset with the corresponding ones of the Baheya dataset to create the new dataset with training, validation, and test subsets, each containing 487, 160, and 163 samples, respectively.
	
	\subsubsection{Brain magnetic resonance imaging (MRI) dataset}
	\label{ssec:Datasets:MRI}
	Ultrasound scans are among the most challenging images for CNNs due to several complexities such as speckle noise, non-uniform resolution, and ambiguous boundaries. To evaluate the generalizability of the study and severity of the shift-variance problem on less challenging and even larger datasets, we employed the LGG (low-grade glioma) segmentation dataset from The Cancer Imaging Archive (TCIA) \cite{Buda2019, Mazurowski2017}, which is comprised of magnetic resonance (MR) images from 110 patients' brain. Image sizes were 256$\times$256 pixels, and the number of slices ranged from 20 to 88. We divided patients into three subsets: training, validation, and test composed of 66, 22, and 22 patients, respectively. Although pre-contrast and post-contrast sequences were also provided, we utilized only fluid-attenuated inversion recovery (FLAIR) with their corresponding manual abnormality segmentation masks. Besides, we only used slices containing at least one abnormality and their corresponding ground truths as two-dimensional images. Finally, our training, validation, and test sets were comprised of 833, 268, 272 samples, respectively. For all datasets, we resampled images and their corresponding masks to an identical size of 138$\times$138 pixels.
	
	Sample images from each dataset can be found in the Supplementary material.
	
	\subsection{Training Strategy}
	\begin{figure}
		\centering
		\includegraphics[width=\linewidth]{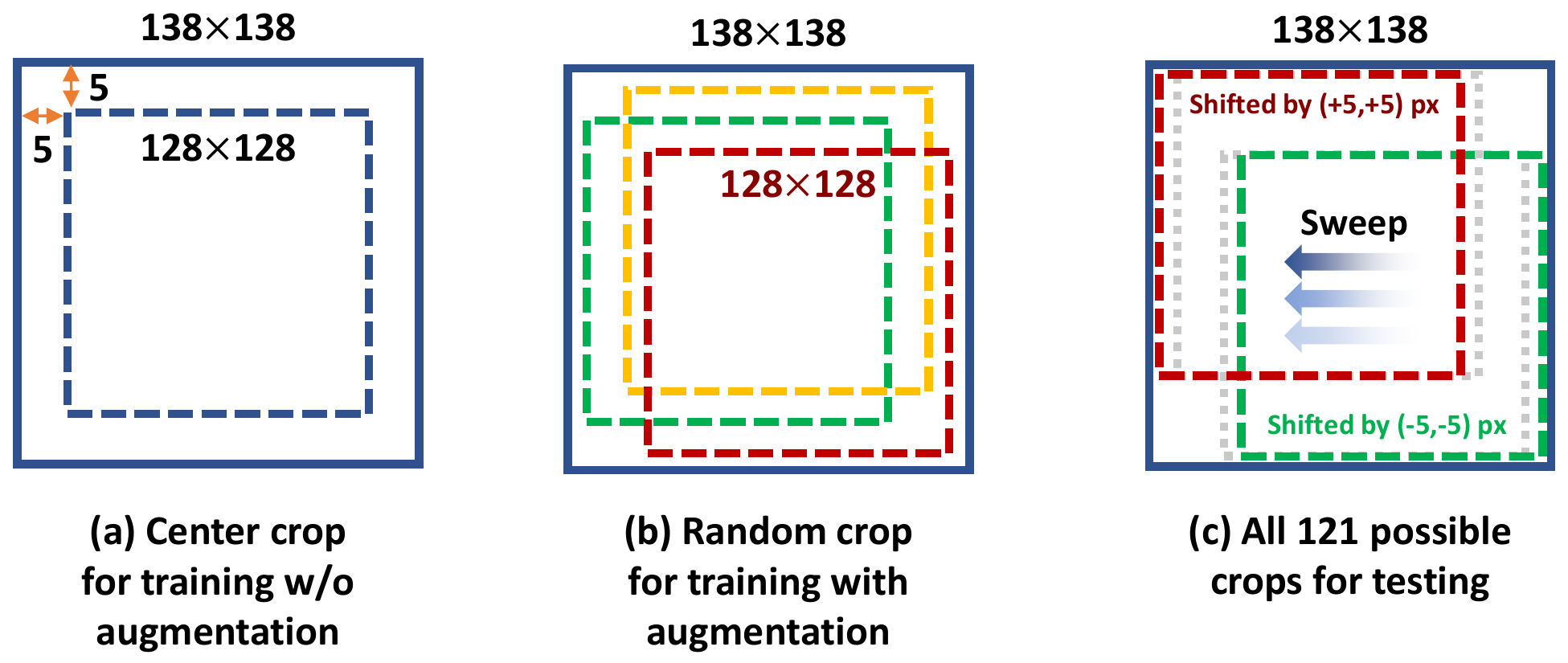}
		\caption{To mimic input translations, we resampled all images to a size of 138$\times$138 in the first stage. (a) For training without data augmentation, where translations were not required, we always center cropped a 128$\times$128 square. (b) For training with data augmentation, at each iteration, a 128$\times$128 square was cropped at a random location. (c) For testing, we mimicked all 121 possible translations. For instance, cropping the top-left region mimics a translated version with respect to the center cropped version where it is shifted by +5 pixels in both horizontal and vertical directions.}
		\label{fig4}
	\end{figure}
	In total, we trained 500 networks from scratch. For each dataset, we trained 5 different networks without data augmentation: a baseline network, three BlurPooling networks with anti-aliasing filters of sizes 3$\times$3, 5$\times$5, and 7$\times$7, and finally a PBP network. Then we trained another 5 networks corresponding to the previous ones and with the same configurations, while this time, data augmentation was applied. Finally, due to small training datasets, we repeated each training 10 times with different random initializations to mitigate the randomness out of the interest of this study. Although those 10 initializations were randomly generated using LeCun method \cite{LeCun2012}, we used 10 fixed initializations and trained the corresponding networks with identical initializations. For instance, if the first repeat of the baseline network (without data augmentation) trained with initialization \#1, the first repeat of the rest 9 networks, i.e., baseline with data augmentation, BlurPooling $m\times m$, and PBP networks (with and without data augmentation) were also initialized with initialization \#1. The same 10 initializations were similarly used across all datasets.
	
	Since DSC is one of the most common metrics for evaluating medical image segmentation, we chose the loss function based on this metric that quantifies the area overlap between the predicted and ground truth masks
	\begin{flalign}
	DSC(\boldsymbol{S},\boldsymbol{\hat{S}}) = \frac{2 \left |\boldsymbol{S} \cap \boldsymbol{\hat{S}} \right |+ \varepsilon}{\left |\boldsymbol{S} \right |+ \left | \boldsymbol{\hat{S}} \right | + \varepsilon}
	\end{flalign}
	\begin{flalign}
	L(\boldsymbol{S},\boldsymbol{\hat{S}}) = 1 - DSC(\boldsymbol{S},\boldsymbol{\hat{S}})
	\label{eq_loss}
	\end{flalign}
	Moreover, the sigmoid function was employed as the activation function of the last layer, and the learning rate and batch size were $2\times 10^{-4}$ and 24, respectively. We utilized AdamW \cite{Loshchilov2019}, a variant of Adam \cite{Kingma2015} as the optimizer, and set the weight decay parameter to $10^{-2}$. AdamW yields a better regularization by decoupling the weight decay from the optimization steps taken with regard to the loss function. To avoid overfitting, in addition to applying a weight regularization strategy, an early stopping strategy was also pursued to stop training when the validation loss stops improving after 100 epochs. For each training epoch, we saved model weights only if the validation loss had been improved and finally used the best weights for testing. We used the same configuration for all experiments. They were programmed using the PyTorch package \cite{Paszke2019}, and training was performed on a single NVIDIA TITAN Xp GPU with 12 GB of memory.
	
	\subsection{Augmentation}
	\label{ssec:Methodology:Augmentation}
	One may wonder that instead of modifying the downsampling layers,  the shift-equivariance in CNNs can be achieved by showing the networks the translated versions of images in the training set. To evaluate the effectiveness of this solution and compare it with the other approach, we conducted each experiment both with and without data augmentation. For the former case, as shown in Fig. \ref{fig4} (a), we trained networks with center cropped images of size 128$\times$128 pixels. For the latter case, since the focus was on achieving shift-equivariance, we augmented the data merely by input translations and isolated experiments from other transformations. To this end, we applied an on-the-fly augmentation by randomly choosing a different square with a size of 128$\times$128 within the original image at each training epoch (Fig. \ref{fig4} (b)). Since the stopping strategy was based on the validation set, we always used center copped images for validation, even for augmented training cases.
	
	To go one step further, we also evaluated the effectiveness of other types of data augmentation methods in achieving shift-equivariance. To this end, in addition to the translation-based augmentation, we randomly scaled, blurred, flipped, and rotated images, adjusted their brightness and contrast, and finally added noise to them. Detailed explanations and results are provided in the Supplementary material.
	
	\subsection{Evaluation Metrics}
	\label{ssec:Methodology:EvaluationMetrics}
	\subsubsection{Accuracy}
	\label{ssec:Methodology:EvaluationMetrics:Accuracy}
	To report the accuracy, we used the segmentation error defined in (\ref{eq_loss}). The error range is [0, 1], and it is closer to 0 for more accurate predictions.
	
	\subsubsection{Consistency}
	\label{ssec:Methodology:EvaluationMetrics:Consistency}
	To quantify shift-equivariance, we evaluated the output consistency across input translations. To this aim, as shown in Fig. \ref{fig4} (c), we mimicked input translations by translating a sample image $\boldsymbol{I}$ from the test set by $i$ pixels horizontally and $j$ pixels vertically with respect to its center cropped version, where $\{i,j \in \mathbb{N} \: |-5 \leq i,j \leq 5\}$. We labeled such a translation as $(i,j)$ and the translated input as $\boldsymbol{I}_{ij}$. Then we obtained corresponding output segmentation masks $\boldsymbol{\hat{S}}_{ij}$ for all 121 possible translated versions. Finally, the variance over segmentation errors was calculated for the test sample $\boldsymbol{I}$, where a lower variance denotes more consistency and higher shift-equivariance.
	\begin{flalign}
	L_{i,j} = L(\boldsymbol{S}_{ij}, \boldsymbol{\hat{S}}_{ij})
	\end{flalign}
	\begin{flalign}
	Error\:Mean\:=\overline{L_{i,j}} = \frac{1}{121}\sum_{i=-5}^{i=5}\sum_{j=-5}^{j=5} L_{i,j}
	\label{eq_errormean}
	\end{flalign}
	\begin{flalign}
	Error\:Variance\: = \frac{1}{120}\sum_{i=-5}^{i=5}\sum_{j=-5}^{j=5}
	(L_{i,j}-\overline{L_{i,j}})^2
	\label{eq_errorvar}
	\end{flalign}
	\vspace{0pt}
	
	\section{Results}
	\label{Results}
	\begin{figure}
		\centering
		\includegraphics[width=\linewidth]{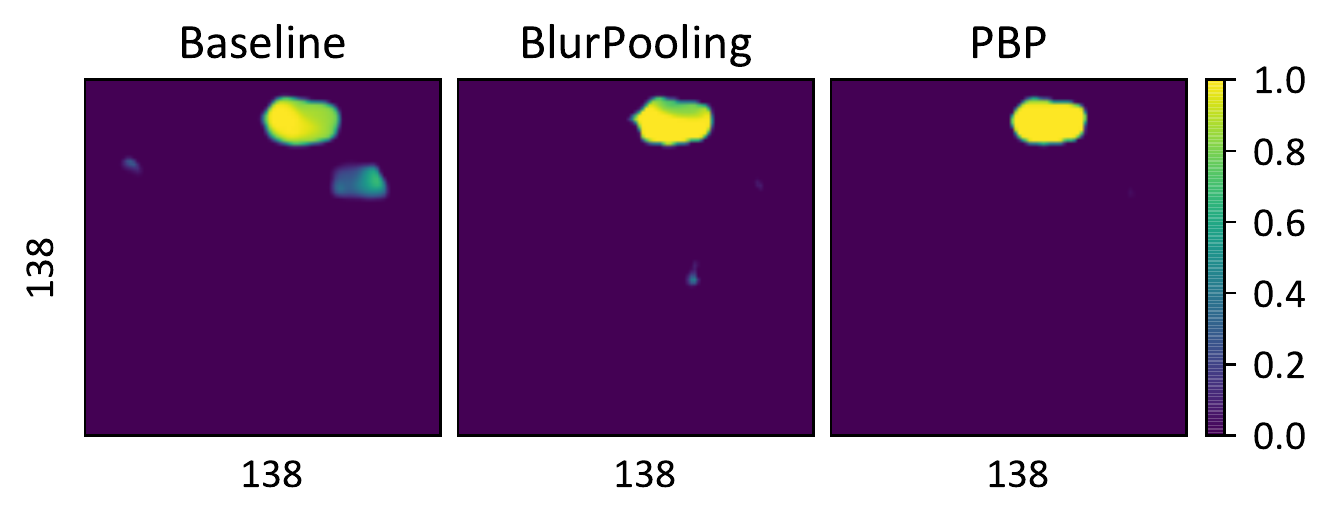}
		\vspace*{-18pt}
		\caption{Illustration of the effect of input translations on generating segmentation masks. An identical input image from the test set was translated by $(i,j)$ pixels, where $\{i,j \in \mathbb{N} \: |-5 \leq i,j \leq 5\}$. Each network generated output segmentation masks corresponding to those 121 translated inputs. Outputs were compensated for translations with respect to the reference and finally averaged over all translations. (Left) Baseline network. (Middle) BlurPooling with an anti-aliasing filter of size $7\times 7$. (Right) PBP U-Net.}
		\label{fig5}
		\vspace{-12pt}
	\end{figure}
	
	\begin{figure}[t!]
		\centering
		\includegraphics[width=\linewidth]{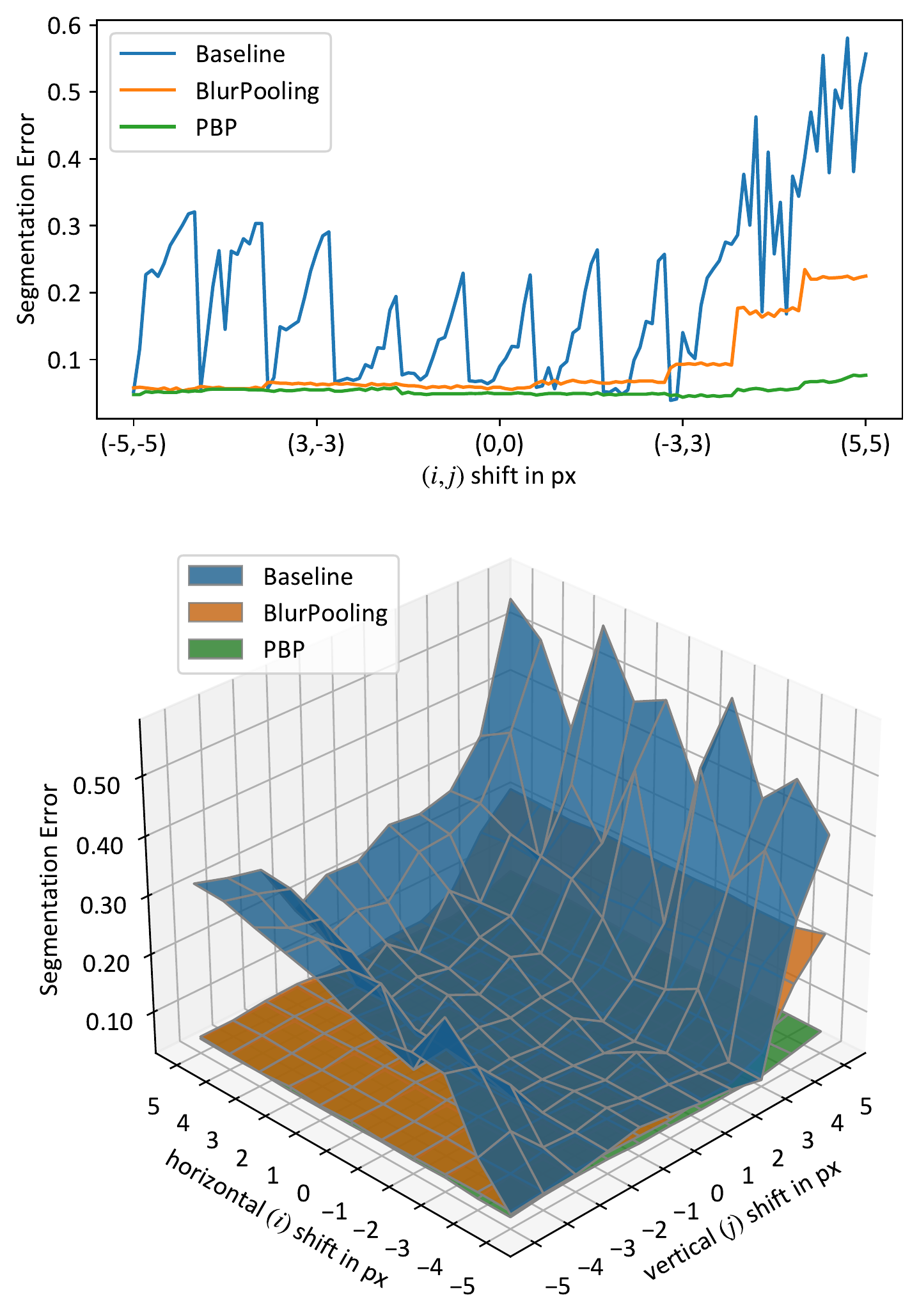}
		\caption{Segmentation errors for 121 translated versions of an identical input image from the test set. The input image was translated for $(i,j)$ pixels, where $\{i,j \in \mathbb{N} \: |-5 \leq i,j \leq 5\}$. (Top) The 2D view wherein $i$ changes faster than $j$ from $-5$ to $+5$. (Bottom) Same values in a 3D view.}
		\label{fig6}
	\end{figure}
	As mentioned before, we trained three types of networks using each dataset: 1) a vanilla U-Net with conventional max-pooling layers referred to as the baseline, 2) three U-Nets, in which max-pooling layers had been replaced with corresponding BlurPooling layers with anti-aliasing filters of sizes 3$\times$3, 5$\times$5, and 7$\times$7, and 3) a U-Net with PBP layers as downsampling layers.
	
	Fig. \ref{fig1} illustrates output segmentation masks corresponding to an identical sample from the test set of UDUAT dataset as an input. The input was translated diagonally for $(k,k)$ pixels, where $k$ $\in \{-3, -1, 0, +1, +3\}$. The left column displays the inputs, and the following columns show the output segmentation masks generated by the baseline method with conventional max-pooling layers, the BlurPooling method with an anti-aliasing filter of size $7\times 7$, and the proposed method with PBP downsampling layers, respectively. In this figure, we limited the results up to 5 translated versions for brevity. For better visualization and to provide a more trustworthy comparison, we generated corresponding outputs for all 121 translated versions of the same test sample. Then we compensated for translations by zero-padding in the output segmentation masks regarding the reference and finally averaged across all translated versions. The results are presented in Fig. \ref{fig5}, where the left, middle, and right images show averaged outputs for baseline, BlurPooling, and PBP methods, respectively. Moreover, the values of segmentation errors corresponding to those 121 translated versions are plotted in Fig. \ref{fig6} in both 2D (top) and 3D (bottom) views.
	
\begin{figure}
		\centering
		\includegraphics[width=\linewidth]{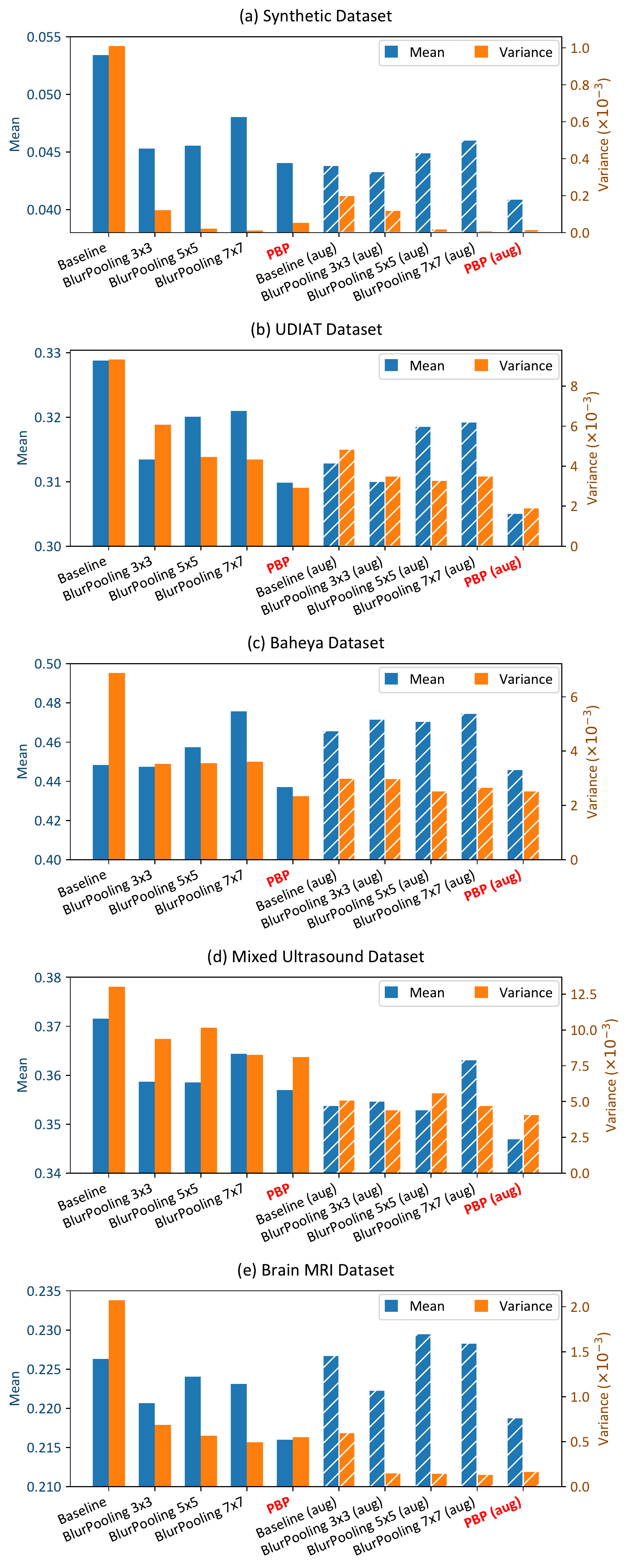}
		\caption{Comparison of segmentation accuracies, as well as output consistencies. 
			Hatched and solid bars represent training networks with and without data augmentation, respectively. The results are demonstrated for (a) the synthetic dataset, (b) the UDIAT dataset, (c) the Baheya dataset, (d) the mixed ultrasound dataset, and (e) the brain MRI dataset. Lower error mean and error variance are better and indicate higher accuracy and consistency, respectively.}
		\label{fig7}
	\end{figure}

	To quantify the output accuracy and output consistency for each method, as mentioned in Section \ref{ssec:Methodology:EvaluationMetrics}, we calculated the mean and variance of the segmentation errors for each test sample $\boldsymbol{I}$ over its all translated versions using (\ref{eq_errormean}) and (\ref{eq_errorvar}). Then we averaged the results over the 10 training repeats with random initializations, and finally over the test set.
	The whole procedure was completed for each method with and without data augmentation during training, and the results for each dataset are illustrated in Fig. \ref{fig7}.

	\section{Discussion}
	\subsection{Consistency}
	\label{ssec:Discussion:Consistency}
	Fig. \ref{fig1} sheds light on the shift-variance problem in CNN-based methods and provides a visual perception of the importance of output segmentation consistency. While diagonally translated inputs in the left column look similar, it can be seen that the baseline method generated substantially different output segmentations in the second column. It demonstrates that translating the input even by one pixel may drastically alter the output segmentation mask. In the third column, applying the BlurPooling method alleviated the problem and improved the output consistency; however, it came at the cost of losing accuracy at the lesion's boundaries. The fourth column demonstrates how the proposed method mitigated the problem and improved the output consistency without compromising accuracy by replacing conventional max-pooling layers with pyramidal BlurPooling layers.  Note that shift values are merely a convention during the test phase, and it is not expected to achieve a lower error for the no-shift case since networks are not aware of the origin.
	Fig. \ref{fig5} gives a broader view of the problem, where output segmentation masks are averaged over 121 translated versions. In addition to detecting wrong regions as the lesion for some translated versions, the baseline method failed to detect the lesion consistently, and the predicted mask at the right side of the lesion is very blurry.
	Conversely, the BlurPooling and proposed methods offered a more robust prediction across translated inputs, where the proposed method achieved a higher accuracy.
	
	Fig. \ref{fig6} confirms the same concept as well, where the error generated by the baseline method fluctuates more intensively over input translations compared to the BlurPooling and the proposed method.
	
	In Fig. \ref{fig7}, it can be observed that applying BlurPooling $m\times m$ layers always reduced error variance (improved consistency) compared to the baseline method. As expected, the general trend is that applying progressively stronger low-pass filters yields higher output consistency. Results for the synthetic, UDIAT, and MRI datasets (without augmentation) confirm that BlurPooling layers with filters of sizes, for instance, $5\times 5$ and $7\times 7$ led to lower error variances in comparison with a $3\times 3$, where we can see that larger filters almost achieved a zero error variance for the simple and ideal synthetic dataset. However, in some cases, such as the Baheya dataset (without augmentation), increasing the filter size did not make a considerable difference in the consistency. Because this dataset was more challenging and the output generated by the network had an intrinsic level of uncertainty. Therefore, some degree of error variance was inevitable, and further improvements were hindered by the saturated consistency, even by increasing the low-pass filter size.
	Results for the mixed ultrasound dataset suggest that combining two datasets from different distributions may lead to a lower output consistency compared to each one of the datasets separately.
	
	\subsection{Accuracy}
	\label{ssec:Discussion:Accuracy}
	Fig. \ref{fig7} shows that for all datasets, except for the Baheya one, utilizing BlurPooling $m\times m$ layers yielded a better accuracy compared to the baseline for non-augmented cases across all $m$ values. It may seem surprising as we generally expect to see degradation in accuracy after applying a low-pass filter. However, as the results demonstrated, it is not always the case, and BlurPooling layers may improve accuracy as well for three reasons: 1) Considering the fact that applying low-pass filters did not add any learnable parameters to the network, they might act as a regularization method and control the capacity of CNN to prevent overfitting. This improvement has been observed in classification applications as well \cite{Zhang2019}. 2) It is worth noting that low-pass filters were not applied on feature maps directly but on densely max-pooled maps instead. As shown in Fig. \ref{fig2} (bottom path), this approach enables the network to take advantage of information that was supposed to be entirely ignored. Therefore, exploiting that information might enhance the accuracy. 3) Although low-pass filters were originally applied to mitigate the shift-variance problem, they may be effective in suppressing spurious noise sources, such as speckle noise or other artifacts in the signal that may make the learning process more challenging. In Fig. \ref{fig7} (a), the results demonstrate that the baseline method with the explained configurations managed to achieve a DSC as high as 94.6\% for the synthetic dataset. Interestingly, applying BlurPooling layers improved the accuracy even in this case because of the aforementioned reasons. For the Baheya dataset in Fig. \ref{fig7} (c), utilizing BlurPooling $3\times 3$ layers could not improve the accuracy in comparison with the baseline; however, it managed to preserve the accuracy while decreased the error variance from  $6.9\times{10}^{-3}$ to $3.5\times{10}^{-3}$, i.e., improved consistency by almost a factor of 2. Comparing the results across BlurPooling $m\times m$ networks shows that accuracy generally decreased by employing a larger filter. It was expected because, at some point, the smoothing effect of the filter compromises its potential advantages, meaning that larger filters may improve consistency at the expense of accuracy.
	
	\subsection{Pyramidal BlurPooling (PBP)}
	\label{ssec:Discussion:PBP}
	Fig. \ref{fig7} also provides a comparison of the proposed method (labeled as PBP) with the baseline and BlurPooling $m\times m$ methods. We demonstrated that the overall trend with increasing filter size was better output consistency and the general trend with decreasing filter size was higher segmentation accuracy. As mentioned in \ref{PBP}, the proposed method employs a pyramidal stack of low-pass filters to combine the best of two worlds, in which the filter size is larger at the first downsampling layer and gradually decreases by moving toward the fourth one, aiming to enhance consistency without compromising the accuracy. As is evident from Fig. \ref{fig7} (b), (c), and (d), for the UDIAT and Baheya, and mixed ultrasound datasets, the proposed method consistently outperformed all baseline and BlurPooling $m\times m$ methods with or without data augmentation, whether in terms of accuracy or consistency. Interestingly, in these cases, the proposed method achieved higher accuracy in comparison with BlurPooling $3\times 3$, as well as better consistency compared to BlurPooling $7\times 7$. As for the former, in the last downsampling layer, low-pass filters were applied on feature maps with a size of 16$\times$16, where most of the energy of the signal concentrated around zero. Therefore, in the pyramidal BlurPooling, we utilized the smallest possible filter size (2$\times$2) to avoid the unnecessary smoothing effect.
	As for the latter, regardless of the accuracy, one might expect that the BlurPooling method with an anti-aliasing filter of size $7\times 7$ must always obtain a higher consistency compared to the PBP method.
	However, accuracy and consistency are not completely decoupled, meaning that changing one of them can affect the other one. For instance, a BlurPooling $7\times 7$ network may hold a high output consistency across most of the input translations; however, in specific circumstances, such as where the lesion is too close to the image borders, the network can fail to detect the lesion for translations toward the borders due to lower accuracy, and consequently, these outliers cause a higher error variance in the final results. Meaning that compare to the BlurPooling $7\times 7$, PBP achieves even more consistency by providing higher accuracy and avoiding those outliers.
	Fig. \ref{fig1}, \ref{fig5}, and \ref{fig6} illustrate this case for a sample image where the BlurPooling $7\times 7$ method ended up with a higher error variance (more fluctuations in segmentation error in Fig. \ref{fig6}) because of failing to detect the lesion accurately for vertical translations, where the lesion was too close to the image border.
	This is also confirmed in Fig. \ref{fig7} (a), and (e), where instead of challenging ultrasound datasets containing lesions with vague boundaries and speckle noise, we applied methods on less challenging synthetic and brain MRI datasets. In these cases, while the proposed method provided higher accuracy compared to BlurPooling $7\times 7$, the consistency is slightly worse in the absence of outliers generated by the inaccuracy of BlurPooling $7\times 7$ in challenging translations.
	
	It is worth noting that although using different sets of filter sizes in PBP can lead to different results, we noticed that choosing a set of low-pass filters with slightly different sizes does not dramatically affect the results, and the concept of not using filters with the same size for all feature maps plays the main role. However, as a rule of thumb, to find the size of the first filter, we visually investigated input images and found the largest possible low-pass filter that can be applied to the input image without noticeably changing the boundaries or visibility of the lesions. In our experiments, the inputs had a size of $128\times 128$, and we chose a $7\times 7$ low-pass filter for the first downsampling layer and decreased the filter size gradually moving toward deeper layers. The same rule can be applied for larger or smaller inputs. If the input size is too small or the network has more downsampling layers leading to very small feature maps at deeper layers, we also have this option to apply low-pass filters only to more shallow layers and leave the deeper layers with conventional max-pooling to achieve better results.
	
	\subsection{Data Augmentation}
	\label{ssec:Discussion:DataAugmentation}
	We reported the results of every experiment with and without data augmentation in Fig. \ref{fig7}, represented by hatched and solid bars, respectively. In most cases, such as the UDIAT, Baheya, and brain MRI datasets, the proposed method without data augmentation outperformed the baseline method even with data augmentation in terms of both accuracy and consistency. However, even in the remaining cases, the proposed method still outperformed the baseline method of the same category (i.e., with or without augmentation). For data augmentation, both its type and parameters need to be chosen carefully concerning the specifications of the dataset. For more illustration, take the classification of images containing handwritten digits as an example, where augmenting the dataset with rotating or flipping may lead to a wrong label for digit "nine" by changing it to digit "six" or vice versa. Despite this evident example, less obvious scenarios can happen as well. In Fig. \ref{fig7} (c) and (e), we can see that utilizing data augmentation increased error means. In the brain MRI dataset, for instance, skull boundaries might be powerful features for the network. Therefore, randomly cropping a portion of boundaries results in a more challenging learning process leading to lower performance. In this case, the crop size can be considered a hyperparameter that needs to be tuned carefully. It is an advantage for the PBP method, which provides a built-in data augmentation without the aforementioned limitation. Moreover, applying BlurPooling layers on top of the data augmentation improved the consistency even further. It demonstrates that data augmentation cannot be considered as a replacement for the proposed method, and they are not interchangeable concepts.
	
	\section{Conclusions}
	\label{sec:Conclusions}
	This study was encouraged by the fact that although accuracy is a principal measure in ultrasound image segmentation using CNNs, the consistency of outputs across different tests must not be overlooked due to multiple clinical motivations. According to the rapidly growing exploitation of CNNs in this domain, and based on what has been reported in the literature, we challenged the common assumption that CNNs are shift-invariant. For the first time, we investigated the shift-variance problem in ultrasound image segmentation or even more broadly in ultrasound images. To cover an extensive range of previous studies, we chose U-Net as the baseline network due to widespread utilization of either its vanilla version or its variants by the community. Moreover, we discussed the origin of the shift-variance problem that enables us to generalize the concept of the study to other networks with different architectures from the U-Net, which still use conventional downsampling layers such as max-pooling or strided-convolutional layers without respecting the Nyquist–Shannon sampling theorem. Demonstrating the existence of the shift-variance problem in the ultrasound image segmentation task, we applied a recently published technique referred to as BlurPooling to mitigate the problem and evaluated its performance with different configurations. For evaluation, we quantified the shift-variance problem using a metric based on error variance and conducted all experiments with and without data augmentation to illustrate that augmentation techniques are not a replacement for modifying downsampling layers. Finally, we presented the Pyramidal BlurPooling method specifically for medical image segmentation, in which the size of blurring kernels decreases gradually at deeper downsampling layers, where more energy of feature maps is concentrated at lower frequencies. Testing on \textit{in-vivo} ultrasound datasets, we demonstrated that the proposed method outperformed the baseline and BlurPooling methods, where it drastically improved the output consistency and, to a lesser extent, segmentation accuracy.
	
	\section*{Acknowledgment}
	The authors would like to thank Natural Sciences and Engineering Research Council of Canada (NSERC) for funding. 
	
	\ifCLASSOPTIONcaptionsoff
	\newpage
	\fi
	
	\bibliographystyle{IEEEtran}
	\bibliography{bibliography}
\end{document}


\title{Supplementary Material for Investigating Shift-Variance of Convolutional Neural Networks in Ultrasound Image Segmentation}
	
	\author{Mostafa~Sharifzadeh,
		Habib~Benali,~\IEEEmembership{Fellow,~IEEE,}
		and~Hassan~Rivaz,~\IEEEmembership{Senior~Member,~IEEE}
		\thanks{M. Sharifzadeh, H. Benali, and H. Rivaz are with the Department
			of Electrical and Computer Engineering, Concordia University, Montreal, QC, Canada, and with PERFORM Center, Concordia University, Montreal, QC, Canada.}
		}
	\maketitle

	\begin{figure}
		\centering
		\includegraphics[width=0.9999\linewidth]{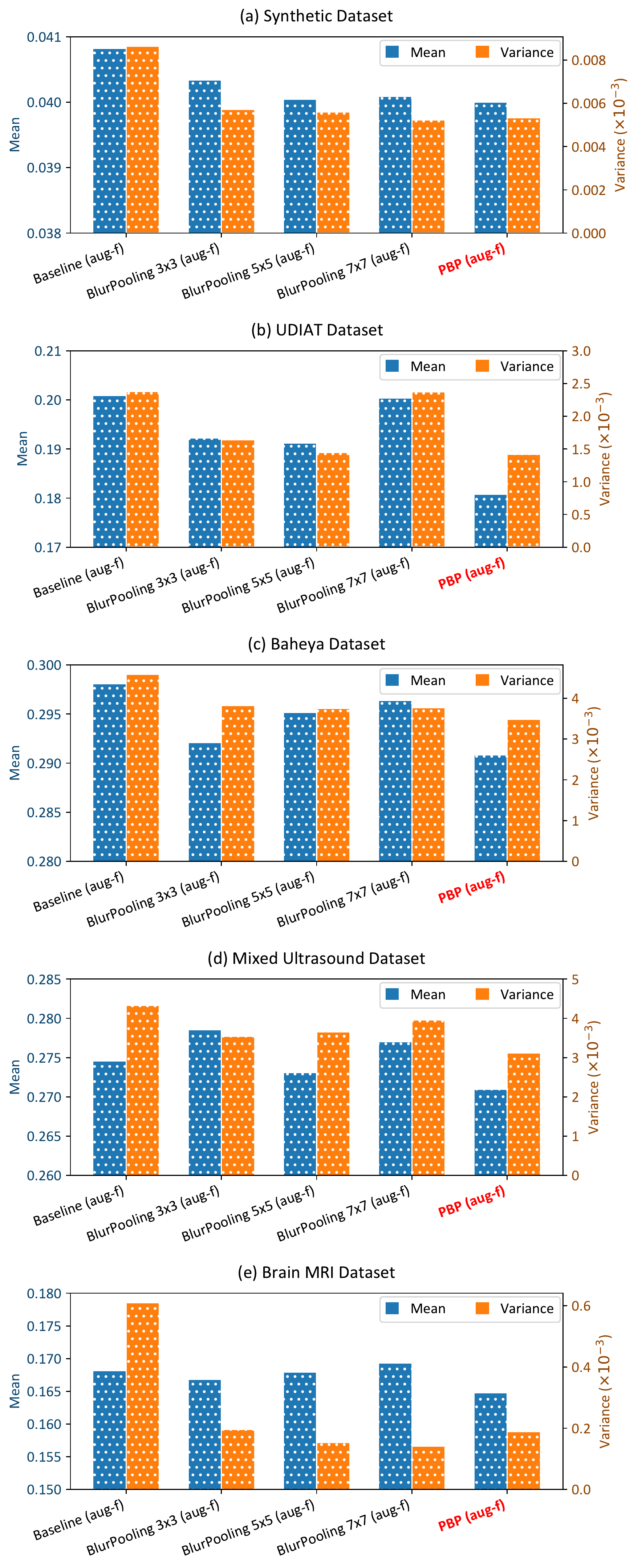}
		\caption{Comparison of segmentation accuracies, as well as output consistencies for experiments with full augmentation (labelled as aug-f). The results are demonstrated for {a) the synthetic dataset, (b) the UDIAT dataset, (c) the Baheya dataset, (d) the mixed ultrasound dataset, and (e)} the brain MRI dataset. Lower error mean and error variance are better and indicate higher accuracy and consistency, respectively.}
		\label{figs-fig7_2}
	\end{figure}

	\section{Full Augmentation}
	\label{ssecs:Full Augmentation}
	Both BlurPooling and proposed methods modify the network's structure to mitigate the shift-variance problem. However, another solution can be showing the networks the translated versions of images in the training set. Therefore, as explained in Section \ref{ssec:Methodology:Augmentation}, we limited the data augmentation only to random translations to concisely compare these two different approaches.
	
	To go one step further and evaluate the effectiveness of other types of data augmentation in achieving shift-equivariance, we repeated all experiments by applying six augmentation techniques. In addition to random translations, we randomly scaled the image by $s\%$ along both axes, where $s\in [-7, +7]$ to make sure that the object will not completely leave the field of view, as well as the $138\times138$ image will not be smaller than $128\times128$. We also altered the image brightness and contrast and applied a Gaussian smoothing filter with a kernel of size $k\times k$ and standard deviation $\sigma$, where $k \in \{0, 2, 3, 5, 7\}$ and $\sigma \in [0, 1]$ were chosen randomly. Besides, we flipped (horizontally) and rotated ($\theta\;degrees$) images, where the chance of flipping was 50\%, and $\theta \in [-10, 10]$. Regarding the fact that ultrasound images are non-stationary along the axial direction and depend on the position of the probe, we did not apply vertical flipping or rotations more than 10 degrees. To avoid creating black regions caused by rotation, we extended image borders by mirroring. Finally, since the speckle noise is the dominant noise source in ultrasound images, and it could affect both output consistency and accuracy by playing an important role in the learning process, we modeled the speckle noise as a multiplicative noise and randomly applied it to the image:
	\begin{flalign}
	\boldsymbol{I_{noisy}}=\boldsymbol{I} + \boldsymbol{NI}
	\end{flalign}
	where $\boldsymbol{N}$ is a matrix with the same size of the image consisting of normally distributed values with zero mean and standard deviation $\sigma \in [0, 0.01]$. A sample set of transformed images has been shown in Fig. \ref{figs-transformations-samples}. 
	
	Since choosing and storing the best model was based on the validation set, we always used center copped images with no augmentation during the validation phase, even for augmented training cases.
	
	Results are illustrated in Fig. \ref{figs-fig7_2}. As we expected, applying more augmentation techniques improved the segmentation accuracy for all datasets. For instance, for the UDIAT dataset, the DSC improved from 67.1\% to 79.9\% after applying full augmentation to the baseline method. We can see that as opposed to the BlurPooling methods, the proposed method consistently maintained or improved the segmentation accuracy compared to the baseline method.
	
	Except for BlurPooling $7\times 7$ applied to the UDIAT dataset, the general trend for output consistency is aligned with the previous results in Fig. \ref{fig7}, wherein applying BlurPooling layers improved this metric. In the aforementioned exception, the low accuracy might cause a slightly worse consistency compared to the baseline method, since as discussed in Section \ref{ssec:Discussion:PBP}, the accuracy and consistency are not completely decoupled. Similar to the segmentation accuracy, the proposed method consistently improved the output consistency compared to the baseline method and maintained or improved output consistency compared to the best BlurPooling method in UDIAT, Baheya, and mixed ultrasound datasets. The proposed method could not obtain a lower output consistency compared to the best BlurPooling method in the synthetic and brain MRI datasets because as discussed in Section \ref{ssec:Discussion:PBP}, these datasets were substantially less challenging than \textit{in-vivo} ultrasound datasets.
	
	As it can be observed, applying the full augmentation technique decreased the difference between baseline, BlurPooling, and proposed methods in terms of output consistency. This originates from the fact that this comparison may not be completely fair, because the evaluation metric was based on translations (explained in Section \ref{ssec:Methodology:EvaluationMetrics:Consistency}), meaning that the output consistency was quantified by translating the input and measuring the error variance across outputs. To calculate the output consistency, if for instance, in addition to merely translating the input, we also add random noise, the output consistency may decrease using the baseline method. This study was focused on investigating the shift-variance problem; however, an area of future work is addressing this concern and providing a fair comparison by quantifying the output consistency using a new metric that takes into account all transformations used for the full data augmentation.
	
	\begin{figure}
		\centering
		\includegraphics[width=0.81\linewidth]{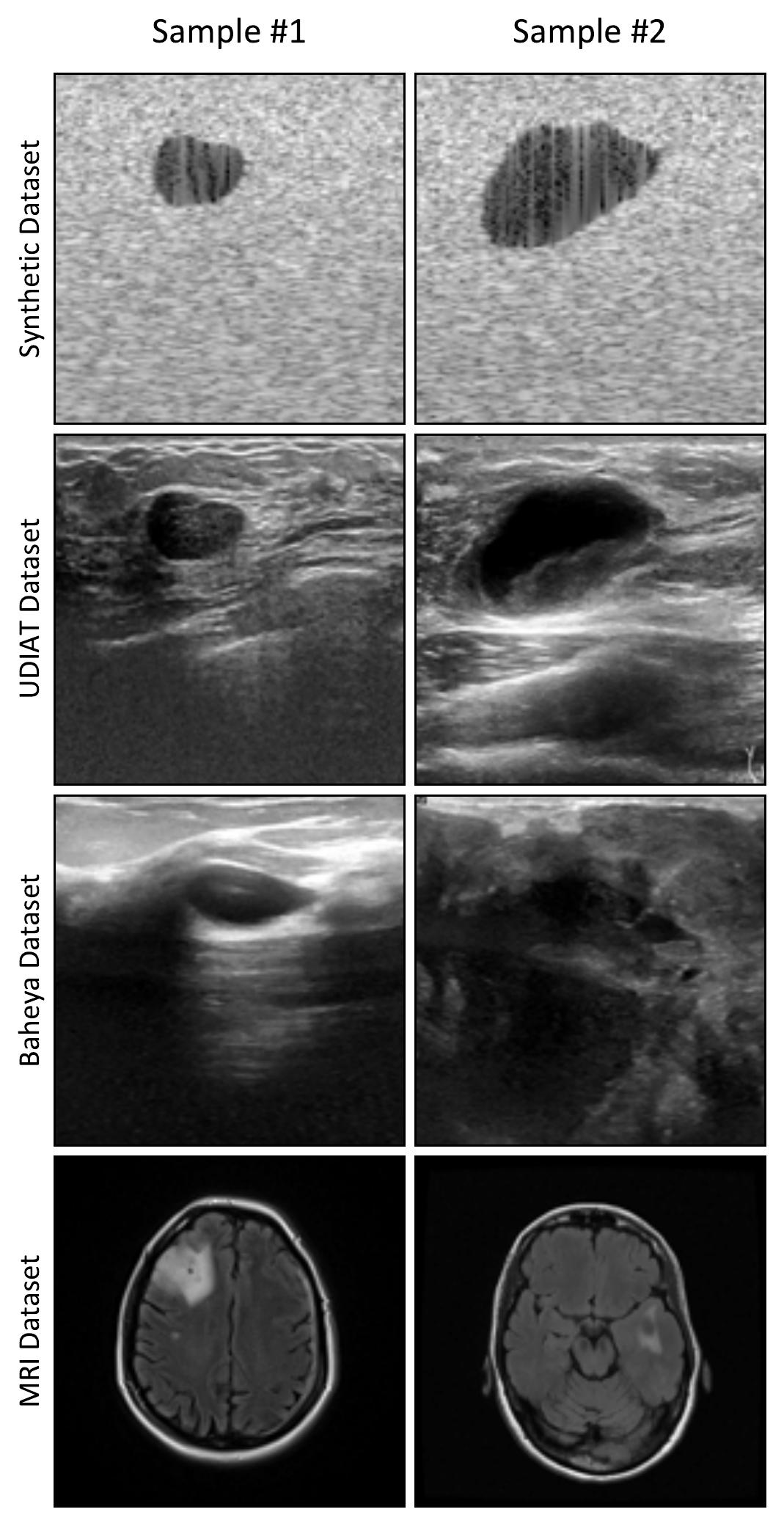}
		\caption{Sample images from datasets employed in experiments. For UDIAT and Baheya datasets, samples \#1 and \#2 belong to benign and malignant categories, respectively.}
		\label{figs-dataset-samples}
	\end{figure}

	\begin{figure}
		\centering
		\includegraphics[width=0.9999\linewidth]{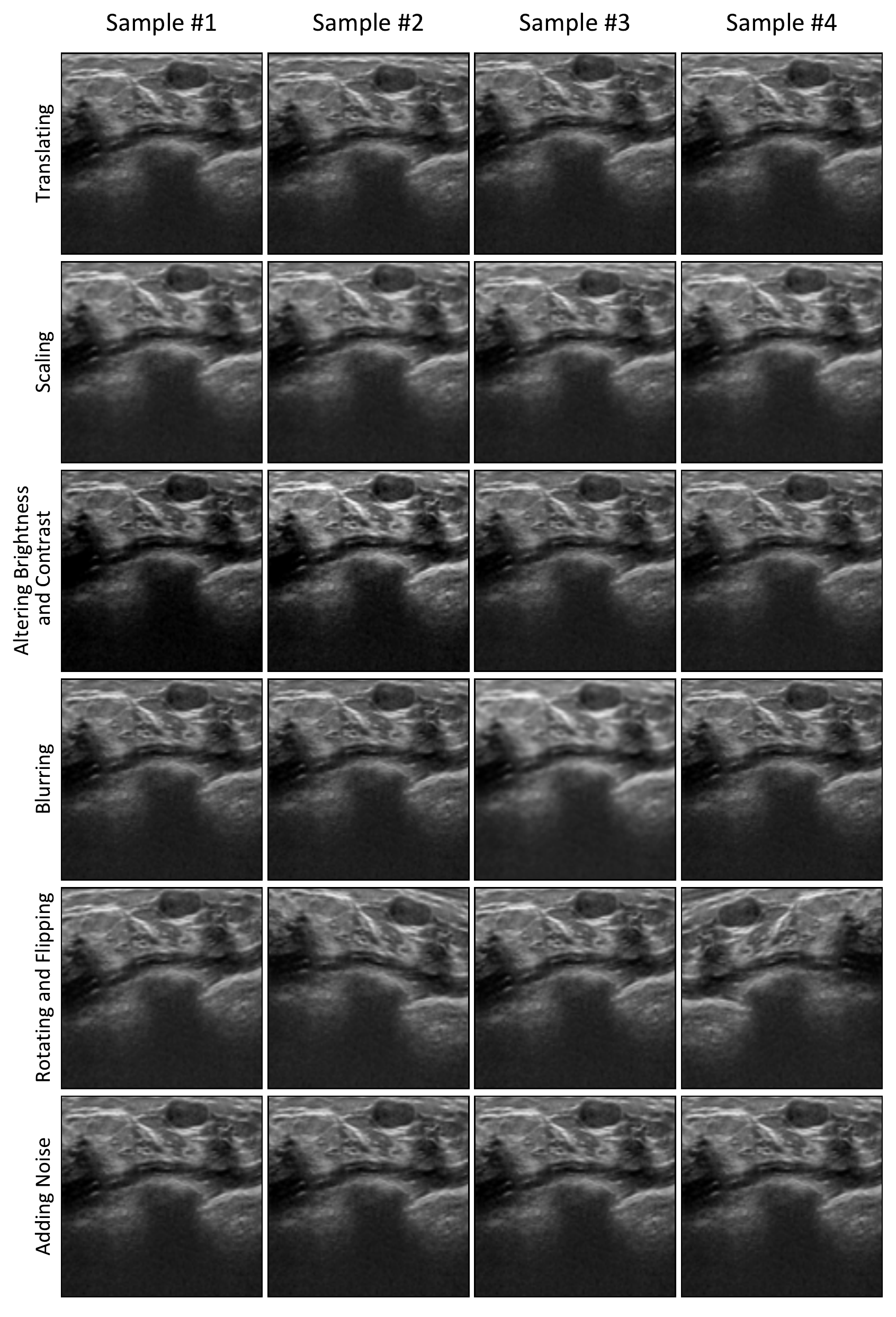}
		\caption{Illustration of transformations applied to input images for the full data augmentation.}
		\label{figs-transformations-samples}
		\vspace{-5pt}
	\end{figure} 

	\section{Datasets Samples}
	Fig. \ref{figs-dataset-samples} shows sample images from datasets employed in this study (Section \ref{ssec:Datasets}), where first, second, third, and fourth rows correspond to the synthetic, UDIAT, Baheya, and MRI datasets, respectively. For UDIAT and Baheya datasets, sample \#1 and sample \#2 belong to benign and malignant categories, respectively.
	
	\section{Errors across Translated Versions}
	\label{ssecs:Errors across Translated Versions}
\begin{figure}
		\centering
		\includegraphics[width=0.9999\linewidth]{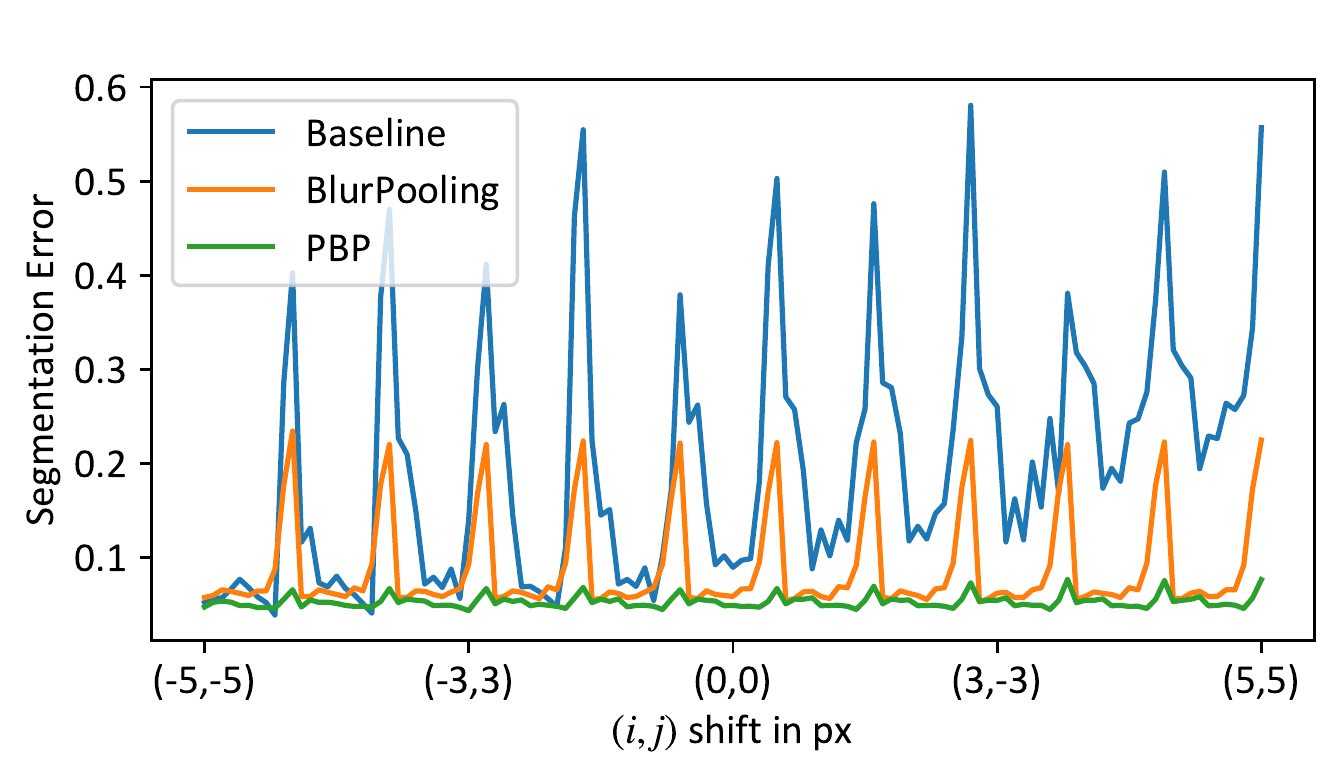}
		\caption{Segmentation errors for 121 translated versions of an identical input image from the test set. The input image was translated for $i$ pixels horizontally and $j$ pixels vertically, where $\{i,j \in \mathbb{N} \: |-5 \leq i,j \leq 5\}$.}
		\label{figs-fig6_2}
	\end{figure}

	Fig. \ref{fig6} (top) illustrates segmentation errors for 121 translated versions of an identical input image from the test set, where the input image was translated for $i$ pixels horizontally and $j$ pixels vertically. To draw this plot, as shown in Fig. \ref{fig4} (c), we started from $j=-5$ and changed $i$ from $-5$ to $+5$. Then we increased $j$ by one and repeated the same procedure to plot all 121 points, meaning that $i$ changes faster than $j$. Without losing generality, we can also plot the values by swapping the priority of $i$ and $j$. In Fig. \ref{figs-fig6_2} same values have been plotted; however, this time the priority of axes has been swapped, and $j$ changes faster than $i$. In this case, we see a periodicity that can be justified by looking at the 3D view of values shown in Fig. \ref{fig6} (bottom).
	
	\ifCLASSOPTIONcaptionsoff
	\newpage
	\fi
	
	\bibliographystyle{IEEEtran}
\makeatletter\@input{xx.tex}\makeatother